Funding: This study received a grant [HI22C1477] from Korea Health Industry Development Institute during 2023 to 2025.

Title: Governance and Technological Challenge in Digital Solidarity Economies: A Case Study of a Collaborative Transportation Platform in South Korea

Author 1: Jeongone Seo, Ph.D, LMSW, Sungkyunkwan University, South Korea

Author 2: Tawfiq Ammari, Ph.D, Rutgers University, U.S.A

Abstract: This paper examines the design, implementation, and eventual shortcomings of a government-led collaborative mobility platform project in South Korea's City P, illustrating how lofty goals of digital solidarity can falter when challenged by local governance realities. Drawing on Hansmann's ownership theory, collaborative governance concepts, and platform cooperativism, we conducted a qualitative case study involving policy documents, independent assessments, and 11 in-depth interviews with residents, officials, and technology developers. Findings reveal a marked disconnect between the initiative's stated emphasis on community co-ownership and the actual power dynamics that largely favored government agencies and external firms. Although blockchain and integrated digital tools were meant to enhance transparency and inclusivity, stakeholders—especially elderly residents—experienced confusion and mistrust. We argue that genuine collaboration in digital solidarity economies requires not only robust technical designs but also culturally resonant ownership structures, substantive inclusion of local voices, and transparent governance mechanisms. The City P case underscores the necessity of addressing heterogeneous digital capacities, aligning funding and incentives with grassroots empowerment, and mitigating performative participation to ensure meaningful, sustainable outcomes in community-based digital innovation.

Teaser: This study reveals how a government-driven mobility platform in South Korea fell short of its cooperative ideals due to imbalances in power, opaque technology implementation, and a lack of genuine community ownership.

Publication fractions: Author 1, Sungkyunkwan University A: 0.700 fraction; Author 2, Rutgers University B: 0.300 fraction.

---

1. Introduction

South Korea's development path has been shaped by dual, contrasting traditions: a strong state-led economic development model and rich grassroots practices of mutual aid. From the 1960s onward, the Korean "developmental state" drove rapid industrialization through top-down planning and close government-business coordination (Amsden, 1989; Kim, 1997). Heavy state intervention and strategic direction were crucial in transforming Korea from a poor agrarian society into an economic powerhouse (Minns, 2001).

At the same time, Korean society retains deep-rooted traditions of collective solidarity and mutual aid at the community level. Historical institutions such as dure, gye, and hyangyak exemplify this legacy. The dure was an age-old farmers' cooperative present in nearly every Korean village by the 18th century, organizing collective labor and mutual assistance through democratic decision-making processes (Choi, 2006). The gye were ubiquitous mutual aid associations functioning as community funds or rotating credit unions managed by their members on a consensus basis (Jung & Rösner, 2012). The hyangyak, or community compact, was a local self-governance agreement originating in the Joseon era that aimed to provide mutual assistance and moral guidance to all members of a village or county (Kim, 2009). These institutions fostered solidarity, with hyangyak compelling residents to support one another and uphold communal norms. In sum, Korea's socio-economic fabric interweaves top-down governance with bottom-up cooperative traditions, setting the stage for contemporary efforts to blend these approaches.

In recent years, the Korean government has sought to leverage advanced technology to rejuvenate local economies by building on these solidarity traditions. In 2017, President Moon Jae-in launched the Presidential Committee on the Fourth Industrial Revolution, known as PCFIR, to promote a "people-centered" agenda for the Fourth Industrial Revolution, referred to as 4IR (PCFIR, 2017). This policy framework envisioned harnessing technologies such as blockchain, Internet of Things, AI, and digital platforms to drive inclusive growth and address regional disparities (Kim & Choi, 2019; UNESCO, 2021). One key goal of the committee was to revitalize underdeveloped regions through collaborative digital platforms, combining high-tech innovation with the ethos of the social and solidarity economy. For example, blockchain-based platforms were envisioned to empower communities to launch new enterprises, such as cooperative multi-modal shared mobility services or e-commerce, thereby stimulating rural economies (Yonhap, 2017).

This vision drew from the idea of the digital solidarity economy, wherein digital tools are deployed in a cooperative, community-oriented manner rather than for purely commercial aims. By infusing traditional mutual aid principles with modern technology, the government hoped to create sustainable local platforms that increase economic participation, improve services, and strengthen community bonds outside the capital region.

However, implementing these ideals has proven challenging. This paper examines a representative case: a government-led collaborative multi-modal shared mobility platform project launched in a provincial city (hereafter "City P") under the Fourth Industrial Revolution (4IR) initiative. The project aimed to create a community-owned service using blockchain and mobile app technology to improve mobility (including shared bicycles, scooters, and automobile sharing) and boost the local economy. Despite high ambitions and significant public support, it ultimately fell short of its goals. This study investigates what factors contributed to the shortcomings of City P's multi-modal shared mobility platform in achieving its stated objectives, and how these factors reflected a misalignment between the theoretical principles of digital solidarity economies and the project's practical implementation.

2. Related Work

In this section, we start by introducing two relevant theoretical frameworks: collaborative governance in §2.1, and enterprise ownership theory in §2.2. We use both to explain how government can act as mediator between different community stakeholders. In §2.3, we discuss earlier literature studying community-centered technology adoption.

2.1 Collaborative Governance

The concept of collaborative governance provides a first lens for analyzing the case. Collaborative governance refers to arrangements in which public agencies engage diverse stakeholders (community members, private actors, civil society) in a collective decision-making process to implement public initiatives. This approach contrasts with both adversarial (conflictual) and managerial (top-down bureaucratic) modes of policymaking. Ansell and Gash (2008) define collaborative governance as "bringing public and private stakeholders together in collective forums… to engage in consensus-oriented decision making." (p.544) Their meta-analysis of 137 cases identified several critical factors for successful collaboration. These include inclusive participation of all relevant stakeholders, balanced power and resource distribution, incentives for genuine engagement, facilitative leadership, and conducive institutional design (clear ground rules, transparency, and forums for dialogue). During the collaborative process itself, key ingredients are face-to-face dialogue, trust-building, commitment to the process, and development of a shared understanding among participants. When done well, collaborative governance can produce a "virtuous cycle" of small wins that deepen trust and commitment, leading to better outcomes (Ansell & Gash, 2008; Chrislip & Larson, 1994).

Applying this theory, one would expect a government-led platform project to actively involve the intended beneficiaries (e.g., local drivers, residents) and other stakeholders (local government, technology providers, possibly existing taxi unions) from the start. The stakeholders should ideally engage in consensus-based planning and problem-solving, with the government acting as a facilitator rather than a controller. Collaborative governance theory highlights that pre-existing social capital (prior cooperation or conflict) will influence the process, and that power imbalances must be managed so no stakeholder dominates (Ansell & Gash, 2008; Imperial, 2005). It also emphasizes the need for continuous communication and trust among stakeholders. In theory, a collaborative platform initiative grounded in these principles would create a sense of joint ownership and mutual accountability, improving its chances of success. Our analysis will use these criteria (inclusion, power-sharing, leadership, trust, etc.) to evaluate how the City P project was governed and where it deviated from an ideal collaborative approach.

2.2 Enterprise Ownership Theory

The second lens is enterprise ownership theory, particularly as articulated by Hansmann (1996) in The Ownership of Enterprise. Hansmann's framework examines how the ownership structure of an organization (i.e., who holds formal control and residual earnings rights) affects its performance and stakeholder alignment. A core insight is that different ownership models (investor-owned firms, worker cooperatives, consumer cooperatives, etc.) involve trade-offs in terms of transaction costs and collective decision-making costs. According to Hansmann, the optimal ownership group for an enterprise is typically the stakeholder class that can minimize the sum of ownership costs (e.g., costs of decision-making, governance among owners) and market contracting costs (the costs non-owner stakeholders would face in arm's-length transactions). For instance, investor-owned firms are prevalent because the costs for dispersed shareholders to own and govern the firm are often lower than the contracting costs that customers or employees would incur. In contrast, cooperatives (enterprises owned by workers or consumers) can be advantageous when they reduce market frictions or asymmetric interest problems, but they may suffer if owners are too heterogeneous in interests, which raises the cost of reaching decisions. Hansmann notes that worker-owned firms are relatively rare partly because workers can have diverse preferences, making consensus difficult and thus increasing governance costs.

For the City P collaborative platform, the intended model was essentially a community-owned multimodal shared mobility cooperative. From an ownership theory perspective, such a model needs to align the rights and incentives of the participant-owners with the goals of the enterprise. If the cooperative is nominally owned by community members but in practice key rights are held by the government or a third-party developer, this misalignment can generate conflict and inefficiency. Enterprise ownership theory directs our attention to questions of ownership rights, governance structure, and profit (or benefit) distribution. Recent extensions of Hansmann's insights to the digital and civic technology domains reinforce this point: for example, Scholz (2016) argues that platform cooperatives can lower transaction and governance costs by giving users and workers ownership stakes, while recent studies of digital government initiatives in both developed (Columbia World Projects, 2021) and developing countries (Peixoto & Sifry, 2017) highlight how the design of ownership and control structures shapes civic platform legitimacy, inclusiveness, and effectiveness. We will examine whether the allocation of ownership and control in the City P platform was conducive to reducing conflicts of interest or whether it instead created new conflicts and costs.

2.3 Digital Solidarity Economy, Platform Cooperativism, and Public Innovation Platforms

Recent academic discourse provides context on digital solidarity economies and platform cooperativism, as well as lessons from similar cases internationally. The notion of a digital solidarity economy refers to the use of digital technologies to advance solidarity economy principles—cooperation, democratic governance, social equity, and mutual aid. This often manifests in initiatives like platform co-ops, community crowdsourcing projects, or open-source collaborations that explicitly aim for broad social benefit rather than investor profit. Platform cooperativism, a term popularized by Scholz (2014, 2016), advocates for platforms (such as ride-hailing, delivery, or online marketplaces) that are owned and governed by their workers or users, in response to the exploitative tendencies of the corporate "sharing economy." Scholz and others argue that such co-ops can democratize the digital economy, giving stakeholders a fair share of value and a voice in decisions. Platform co-ops are built on the long-standing cooperative principles (voluntary and open membership, democratic member control, member economic participation, etc.), but they face new challenges when implemented via mobile apps and online.

There is also a body of work examining government-led innovation platforms and social economy initiatives (Kim & Choi, 2019; OECD, 2023; Scholz & Schneider, 2016). Governments in various countries have experimented with creating cooperative or public-oriented platforms as an antidote to monopolistic tech giants. Internationally, some success stories exist: for instance, in Barcelona, the city supported platform co-ops and data commons as part of its technological sovereignty agenda, giving rise to projects like Decidim (an open-source participatory democracy platform) which empowered citizens in decision-making (Scholz & Schneider, 2016). In Canada, the co-op ride-hailing service Eva in Montreal is often cited as a promising example of a platform cooperative challenging Uber's model, backed by supportive local regulations. Failures are also instructive: early attempts at platform cooperatives in Europe (e.g., some cooperatively owned food delivery apps) struggled to attain the network effects enjoyed by venture-funded competitors and often faced internal governance hurdles (Rafélis de Broves, 2022).

In South Korea's domestic context, the landscape of platform experiments has been dynamic. The country's rapid uptake of platforms like KakaoTaxi (the near-monopolistic ride-hailing app) and Coupang (e-commerce) has spurred debates on platform regulation and alternatives. The government has occasionally intervened – for example, banning Uber's operations at one point and later pushing for a national taxi platform – but these top-down interventions have had mixed outcomes. There are thousands of cooperatives and social enterprises in Korea (especially since the Framework Act on Cooperatives of 2012), yet "platform cooperatives" remain nascent (Kim & Choi, 2019; Ji, 2018).

Overall, the literature suggests that to succeed, digital solidarity economy projects must reconcile participatory governance with technical and business viability. They should proactively address challenges such as member engagement, conflict resolution, sustainable funding, and competition with dominant players. Government-led projects bring additional complexity: the state can provide resources and legitimacy, but if its approach is too top-down or politically driven, it might conflict with the grassroots empowerment central to solidarity economy principles. In the case examined here, the original policy goal was to establish a government-supported cooperative platform, but the project never evolved into a robust community-owned structure. The following sections detail the context, research approach, and findings across five analytical dimensions that emerged from combining collaborative governance and ownership theory lenses.

3. Research Methods

In this section, we start by introducing our case study in §3.1. We then describe the data collection process (in §3.2), and finally introduce the analytical framework used in our qualitative analysis in §3.3.

## 3.1 Case Background: The City P Multi-Modal Shared Mobility Platform

This study critically examines a government-led collaborative multi-modal shared mobility platform project implemented in City P—a mid-sized urban center in a less developed region. City P faced typical urban challenges, including chronic traffic congestion, insufficient parking infrastructure, and the growing need for environmentally friendly transportation alternatives. In response, local authorities designed an integrated shared mobility platform that encompassed services such as shared bicycles, shared kick scooters, shared parking, and shared automobiles.

The original policy discourse emphasized establishing a community-owned enterprise—recognized as a type of Special Purpose Company (SPC) under Korean Ministry of the Interior and Safety guidelines (2021)—where local members would own and operate the platform. Nevertheless, governance gradually became dominated by government agencies and external developers. Initial funding and technical support came from the central government, and a consortium of private technology firms and urban planning experts took charge of development. The system included novel features like blockchain-based transaction records to enhance transparency.

Multiple obstacles, however, prevented the realization of key objectives. These challenges included opposition from existing transportation stakeholders, inconsistent expectations across different groups, and limited technological adoption in the broader community. Varying degrees of involvement—from eager local champions to passive residents—further destabilized the project's collaborative aspirations.

To explore how the initiative's theoretical design diverged from its practical outcomes, this study applies qualitative research methods. We conducted semi-structured interviews with local residents, implementers, and municipal officials, then performed a thematic analysis focusing on governance processes, technology design choices, and broader socio-political dynamics.

## 3.2 Data Collection

This research employs a qualitative case study approach. Multiple sources of data were collected to capture different perspectives on why the project failed to achieve its goals. First, we obtained City P's project proposal and planning documents, meeting minutes from the platform's steering committee, and policy briefs from the central government outlining the goals of the Fourth Industrial Revolution regional initiatives. For anonymity and consistency, we refer to these internal documents—particularly the initial proposal and associated planning materials—as the internal project proposal document (2019) throughout this paper. These materials provided insight into the project's intended design, objectives, and official narratives. Secondly, when the project encountered difficulties, City P's municipal council commissioned an independent assessment to diagnose key issues (especially after public disputes arose among stakeholders). We reviewed this assessment report, which recorded conflicts (e.g., between community-minded participants and the app developer, between incumbent transportation interests and local drivers) and offered recommendations. Local press coverage and city council proceedings also served as sources, forming a timeline of key events like app launch delays, driver protests, and the eventual service suspension.

We also conducted interviews with 11 individuals directly involved in or knowledgeable about the project (see table below). Nine interviewees consented to audio recording and verbatim transcription, whereas two refused recording; for these latter two, we relied on detailed note-taking. The interviewees included community leaders, project managers, city officials, tech firm staff, and resident/gig workers, with perspectives both for and against the initiative. One of the lead researchers had also previously directed the municipal council's conflict assessment, thereby securing access to critical stakeholders. Table 1 below shows the details of our respondents.

Our research team also engaged in direct observations during the pilot stage. One researcher attended two public briefings organized by the initiative's steering group (as an observer) and later witnessed a dissolution meeting where members debated the project's future. Field notes documented participant attitudes, the atmosphere of discussions, and the degree of trust or frustration exhibited.

This study adhered to strict research ethics. Approval was obtained from the Institutional Review Board (IRB) at Sungkyunkwan University, specifically for the secondary use of the previously collected interview data and other materials. All data were anonymized per IRB guidelines to maintain confidentiality.

Table 1. Participants Information

| Interviewee | Sex | Age | Role |
| --- | --- | --- | --- |
| P1 | M | 40s | Local project manager |
| P2 | F | 60s | Resident/Supporter |
| P3 | F | 50s | Local leader of Clan village/Supporter |
| P4 | F | 30s | Tech firm local staff |
| P5 | M | 40s | City Official |
| P6 | M | 40s | City Official |
| P7 | F | 70s | Resident/gig worker |
| P8 | M | 70s | Resident/gig worker |
| P9 | M | 50s | Resident |
| P10 | F | 60s | Resident/opponents |
| P11 | M | 30s | Tech firm/property owner |

By triangulating these different sources, we formed a multi-faceted understanding of the project's trajectory. We analyzed the data thematically, using codes aligned with our theoretical lenses—collaborative governance and enterprise ownership— to probe issues such as stakeholder engagement, leadership, technology use, and trust/conflict. When presenting these findings, we quote participants anonymously and summarize documented evidence in a way that preserves confidentiality (Braun & Clarke, 2014).

3.3 Analytical Framework

We examine the case through five analytical dimensions derived from combining collaborative governance theory with enterprise ownership theory. These dimensions guided both our coding of data and the structure of our results:

Stakeholder Inclusion and Participation: Assesses who participated in the platform's governance and development, and whether these participants had meaningful roles. This aligns with the collaborative governance emphasis on broad and inclusive stakeholder engagement.

Ownership and Control Rights: Investigates formal and informal ownership allocation—who legally owned the platform and who ultimately directed key decisions. Using enterprise ownership theory, we explore how effectively the cooperative model was implemented and whether any actor dominated control.

Leadership and Facilitation: Focuses on how leaders—local government officials and community champions—fostered or hindered collaboration. Drawing from collaborative governance, we ask whether leadership was facilitative or directive, and whether strong champions emerged to mediate conflicts.

Resource Contribution and Sustainable Incentives: Examines the financial, technical, and organizational resources committed (funding, technology, expertise) as well as the incentive structures in place. We assess whether participants had sufficient rewards relative to their costs and evaluate how well technological choices, like the blockchain-based platform, fit local needs. We also consider capacity-building measures and their adequacy for long-term sustainability.

Trust, Transparency, and Mutual Accountability: Analyzes relational dynamics—whether stakeholders built trust in each other and in the platform's processes. This dimension includes looking at transparency mechanisms (e.g., blockchain, public communication) and the management of conflicts. It probes whether the initiative ultimately fostered a sense of community responsibility or was marred by growing mistrust.

The subsequent Findings section (Section 4) is organized around these five dimensions, contrasting the project's plans with its actual outcomes and illuminating the key factors behind the initiative's difficulties.

4. Findings

We organize this section based on the analytical framework described in §3.3. Each subsection shows how our findings reflect on current theoretical frameworks. Despite the project's ambitious vision of participatory governance, community co-ownership, and inclusive digital innovation, our findings reveal a consistent pattern of disconnect between planned intentions and actual implementation. Residents were often unaware of the initiative's goals, excluded from meaningful decision-making, or overwhelmed by unfamiliar technologies. Leadership was fragmented, transparency mechanisms failed to gain trust, and the promised cooperative infrastructure never materialized. Instead of fostering solidarity and local stewardship, the platform reproduced mistrust, technical opacity, and political fragmentation—ultimately undermining its own sustainability. These findings point not to partial shortcomings, but to a structural failure in enacting the project's core democratic ideals.

4.1 Stakeholder Inclusion and Participation

Planned Vision

The original proposal placed residents at the forefront of designing and operating the initiative. A community-owned enterprise (recognized under SPC law) was expected to manage the platform, with formal programs to enhance citizen capacity and ensure a wide range of voices shaped the services (Internal project proposal document, 2019).

Execution on the Ground

However, on the ground, we found that involvement was confined mostly to a handful of motivated activists, while many local residents seemed uninterested. P2 described how, "At first I was like, 'What on earth is this platform project?' That's when I started to find out about it." This was also echoed by P3, a community leader, who commented, "Why should I have to pay to park right outside my own house?" This highlights how most residents were unaware of the initiative's aims until it was already underway. P2 further remarked, "Whether we do it or not, it's all the same." This sense of apathy about the project was also reflected by gig workers like P6. Moreover, according to P1, a local community manager, digital outreach—via web or app notifications—proved ineffective for older community members who were not accustomed to smartphone-based communication.

Community Division and Weakening Solidarity

Lingering tensions from past redevelopment debates further splintered the neighborhood. Rumors circulated regarding who might gain financially. One particularly contentious policy was newly designated parking lines, prompting complaints from P3 and P2, saying "Why should I have to pay to park right outside my own house? The city painted a parking line in front of my place and then charged me for using it." Such grievances deepened distrust, dissuading broader participation.

4.2 Ownership and Control Rights

Planned Structure

From the outset, project documents described a Special Purpose Company (SPC) that would coordinate government entities, private firms, and community members under a shared-ownership model (Internal project proposal document, 2019). Any revenue generated by new mobility services would theoretically flow back into local pockets, ensuring equitable returns.

Reality of Implementation

Contrary to the plan for collective ownership, control over both development and day-to-day operations remained largely in the hands of government offices and external tech companies. Many residents felt excluded from decision-making processes. According to P1, a project manager, one former tech company employee who later left the project described the entire venture as "a fraud," claiming that no clear information had been provided regarding revenue management, user data handling, or intellectual property rights. This sentiment was echoed by other technological stakeholders like P6. This disparity between stated cooperative ideals and actual corporate-public dominance led to frustration and confusion within the community.

4.3 Leadership and Facilitation

Planned Governance Structure

The proposal envisioned a collaborative governance approach, with local officials offering facilitation rather than top-down directives and dedicated community leaders serving as champions (Internal project proposal document, 2019). The SPC framework was supposed to be resident-driven, supporting open dialogue to refine the platform's design.

On-the-Ground Dynamics

In practice, consistent leadership failed to take hold. While some municipal officials tried to mediate among firms, residents, and civic organizations, deep-seated distrust and conflicting interests ultimately undermined these efforts. Prominent local advocates, including P3 and P6, became the subjects of rumors and skepticism. As one P3 reflected, accusing a local leader of monopolizing resources and information, "[the] director [is] taking everything for herself." Exhausted by such allegations, several key organizers eventually withdrew, creating coordination gaps that further weakened collaborative efforts.

4.4 Resource Contribution and Sustainable Incentives

Planned Resource and Technology Framework

The proposal featured a robust technological foundation, including blockchain-based DID,

Edge AI, and an integrated cloud platform (Micro-MaaS). User-centered incentives—such as point systems for car sharing and parking—were meant to stimulate participation while generating revenue (Internal project proposal document, 2019). The idea was to align economic rewards with community benefit and thus sustain the enterprise over time.

Actual Execution

On the ground, many elderly residents struggled with the necessary mobile applications. For example, P2 said that "even if you install the app, we have no idea how to use it. We end up asking our grandchildren." P3 and P6 shared similar negative experiences as they tried to engage with new technology. Although training sessions were offered, they proved insufficient to bridge the digital divide.

Distrust in the Implementing Firm

Questions also emerged regarding whether the contracted firm actually had the technical expertise it claimed. As P2 noted, "I wasn't sure if the company actually had the technical skills. They just kept saying they'd do an assessment." This skepticism reflected a broader pattern in which much of the promised technological innovation failed to materialize. P6 further described how key system deliverables were either delayed or left incomplete, fueling doubts about the overall feasibility of the project. In addition, system integration and maintenance fell behind schedule, leaving critical features underused or entirely non-functional.

4.5 Trust, Transparency, and Mutual Accountability

Planned Mechanisms for Building Trust

The proposal heavily emphasized blockchain for transparent data sharing, along with public forums and local fairs to encourage open communication (Internal project proposal document, 2019). These measures aimed to preempt conflicts and build long-term confidence in the initiative.

Failure to Achieve Transparency

In practice, blockchain remained an unfamiliar "black box" to most participants. For instance, P2 asked, "What blockchain are they even talking about? I couldn't see anything," highlighting how the complexity and opacity of certain technological systems can undermine collaborative governance by creating user uncertainty. Building on this point, P6 emphasized that insufficient disclosure about who managed the parking lines or controlled user data fueled rumors of corruption and hidden agendas. Complaints eventually escalated into accusations of misconduct, as frustrations over opaque decision-making grew. Instead of resolving tensions, many participants withdrew—according to P1 (local project manager), P3 (local leader), and P5–P6 (city officials)—leaving behind unresolved grievances and a fragmented sense of communal responsibility.

In addition to technical opacity, several participants voiced concerns that the initiative prioritized political optics over genuine collaboration. P1, for instance, criticized the government's focus on highly visible outputs—such as freshly painted parking lines and staged public events—while deeper structural needs, including digital literacy, sustainable governance, and conflict resolution, were overlooked. Similarly, P6 and P7 observed that these performative acts reinforced the perception that final authority remained with state actors rather than the community itself, thereby undermining the project's stated commitment to bottom-up governance.

5. Discussion

In this section, we will present how our findings contribute to current literature in §5.1. We then discuss socio-technical design recommendations in §5.2. Finally, in §5.3, we show how our findings have policy implications for future technological implementations of digital solidarity platforms in the context of collaborative governance.

5.1 Ownership, Collaboration, and Cooperative Platforms

While our findings affirm key insights from existing research on ownership structures, collaborative governance, and platform cooperativism, they also extend these frameworks in important ways. Through the City P case, we identify conceptual gaps in how prior literature addresses the interplay of culture, governance, and technology in digital solidarity initiatives.

Extending Ownership Theory: Cultural Legitimacy as a Third Dimension

Hansmann's (1996) ownership theory emphasizes minimizing governance and contracting costs when determining optimal ownership structures. However, our findings suggest that for digital solidarity platforms, cost minimization alone is insufficient. In contexts like City P, community members' willingness to participate and trust new platforms also hinges on cultural legitimacy, that is, the extent to which ownership models resonate with longstanding traditions of mutual aid such as the dure, gye, and hyangyak systems. We propose extending Hansmann's framework to recognize cultural fit as a third critical factor shaping ownership viability in community-based digital enterprises.

Extending Collaborative Governance Theory: The Risks of Symbolic Inclusion

Collaborative governance theory (Ansell & Gash, 2008) stresses inclusive participation, trust-building, and shared decision-making as foundations for successful initiatives. Yet our findings reveal that merely inviting stakeholders to participate—without granting them meaningful authority—can erode trust rather than build it. In City P, residents who perceived their involvement as symbolic or superficial became disillusioned and increasingly distrustful of the process. We thus refine collaborative governance theory by highlighting the dangers of "performative participation," suggesting that power-sharing must be substantive to sustain stakeholder commitment over time.

Extending Platform Cooperativism: The Complex Role of State Sponsorship

Scholz's (2016) platform cooperativism advocates for worker- and user-owned digital platforms as democratic alternatives to corporate models. However, City P illustrates that when governments act as primary sponsors of cooperative platforms, they may inadvertently reproduce hierarchical structures they intend to dismantle. Without clear institutional safeguards, public sector actors can dominate decision-making, turning nominally cooperative projects into de facto public–private enterprises. Our findings suggest that future theorizing about platform cooperativism must grapple more directly with the tensions introduced by state involvement, particularly around autonomy, accountability, and ownership control.

5.2 Sociotechnical Adoption Implications: Digital Solidarity in Complex Stakeholder Contexts

Technology played a pivotal but ultimately underappreciated role in the City P platform's shortcomings. As emphasized in platform cooperativism literature (Scholz, 2016) and civic tech studies (Columbia World Projects, 2021), technical usability is fundamental for sustaining community engagement. In City P, the mobility app's unreliability, poor user support, and opaque blockchain integration severely eroded citizen trust. Despite aspirations of transparency and empowerment, the platform remained confusing and inaccessible, particularly to older residents with limited digital literacy.

However, beyond these technical failures, our findings highlight a deeper sociotechnical misalignment. Effective digital solidarity platforms must not only meet high usability standards but also embody governance and participation practices that resonate with local traditions of collective action. In the Korean context, this means designing systems that echo the participatory ethos of dure, gye, and hyangyak institutions—systems historically grounded in transparency, mutual accountability, and inclusive decision-making. When technology platforms impose unfamiliar, opaque governance models, even sophisticated tools can alienate intended users.

Moreover, the attempt to merge blockchain with collaborative governance illustrates a critical lesson: technologies intended to enhance trust must be designed in ways that are socially legible. While blockchain was featured prominently in official documents, it remains unclear whether such high-end technology was necessary or appropriate for addressing the needs of a digitally marginalized, elderly community. Its inclusion appeared to serve more as a signal of technological sophistication—useful for attracting government support—than as a solution grounded in local realities. If users cannot understand or verify how blockchain mechanisms protect their interests, technical sophistication becomes irrelevant or even counterproductive. Visibility, simplicity, and community education are essential components for embedding trust in technologically mediated solidarity economies.

Finally, our findings suggest that digital solidarity initiatives must explicitly address heterogeneous digital capacities from the outset. Universal smartphone adoption cannot be assumed. Strategies such as multi-modal access points (e.g., web, app, SMS), localized digital literacy programs, and peer-supported onboarding processes are necessary to ensure broad inclusion. Without these measures, projects risk reproducing new forms of digital exclusion, undermining their social mission.

Taken together, these sociotechnical lessons underscore that technological design must align not only with functional standards but also with the cultural, organizational, and trust-building imperatives of the communities such platforms aim to serve.

5.3 From Dure to Digital: Strengthening Local Ownership in Platform Governance

The City P experience highlights the need for governments to adopt a "partner state" approach (Scholz & Schneider, 2016), supporting but not dominating digital solidarity initiatives. Collaborative governance theory (Ansell & Gash, 2008) emphasizes that authentic empowerment requires the deliberate distribution of power, iterative consensus-building, and visible trust-building milestones.

In the Korean context specifically, future initiatives would benefit from grounding digital solidarity projects in long-standing indigenous traditions of mutual aid and collective governance, such as the dure (agricultural labor cooperatives), gye (rotating credit associations), and hyangyak (local self-governance compacts). These institutions historically fostered strong community bonds through democratic participation, consensus-driven decision-making, and mutual accountability (Kwak, 2012; Jung & Rösner, 2012; Kim, 2009). Rather than imposing externally designed governance frameworks, policymakers should engage local communities by explicitly drawing on these cooperative legacies, adapting their principles to contemporary digital platforms. Doing so can strengthen cultural legitimacy, deepen social capital, and provide familiar models of collaborative practice that are better suited to local expectations.

Policy design should therefore prioritize decentralized decision-making structures, transparent resource governance, and mechanisms to sustain trust, particularly where historical tensions or weak social capital exist. Funding models must encourage gradual scaling—starting with small, achievable pilots—and incorporate robust technical support to meet the usability expectations shaped by private sector standards (OECD, 2023).

Finally, as Hansmann (1996) cautions, attention must be paid to ownership structures. Without careful management of governance costs and stakeholder heterogeneity, community platforms risk replicating the very hierarchies they were meant to displace. Sustainable digital solidarity economies thus require a fine balance between state facilitation, technological excellence, and genuine stakeholder co-ownership, rooted not only in international best practices but also in Korea's rich traditions of local self-organization.

## 6. Conclusion

The ultimate lesson from the City P project is that fostering a locally governed, cooperative digital platform demands more than strong rhetoric or a supportive legal framework. Although the concept of a community-owned mobility service aligned well with Korea's history of grassroots mutual aid and with contemporary policy visions of inclusive innovation, the practical reality faltered. Collaborative governance theory and platform cooperativism both assume that all relevant stakeholders—from government agencies to ordinary residents—will engage in authentic, continuous dialogue and decision-making. Hansmann's perspective reminds us that this engagement requires stable mechanisms to keep collective decision costs manageable, ensure transparent ownership rights, and align members' incentives. None of these conditions were fully satisfied in City P.

Several policy implications follow. First, governments might do better to adopt a facilitative "partner state" stance, granting communities the time and space to learn, deliberate, and build capacities for leadership. Imposing strict deadlines or focusing on immediate metrics often conflicts with the long-term goal of embedding democratic control. Funding mechanisms can be structured to encourage iterative growth, so that after a small-scale pilot demonstrates feasibility, the project can gradually expand. Next, community-oriented digital platforms are inherently complex, demanding competitive technical features if they are to attract users in a market saturated by large private apps. Governments could direct specialized grants or technical support toward cooperative platforms so that they can develop user-friendly software, robust data systems, and effective marketing strategies. Doing so would acknowledge the fundamental reality that technology-driven services require expertise and resources that small cooperatives may struggle to obtain on their own.

When new technologies like blockchain are introduced, they cannot remain abstract promises of transparency. They must be clearly communicated and integrated in ways that bring tangible benefits: for example, verifying transactions or distributing user incentives in a way that is easily visible and verifiable to all participants. If technology remains a "black box," then trust, which is already fragile, deteriorates further. Equally important is addressing digital inclusion from the start. Phone-based reservations or on-site kiosks might be essential for older citizens who cannot use an app. Regular training workshops, possibly run by local youth or civic groups, could bridge generational divides. This is not simply a matter of courtesy but a core aspect of ensuring that the platform achieves enough participation to remain economically and socially viable.

Design lessons also arise from the need to produce visible success stories that can counter skepticism and unify stakeholders around small but meaningful achievements. Rather than attempting an immediate city-wide launch of every planned feature, focusing on a particular neighborhood or specific function can generate early positive feedback. Once trust in that narrower pilot is established, the project can scale in a more stable way. At each stage, governance must be transparent. Meeting records, decision rationales, and financial details should be accessible so that rumors and mistrust do not flourish. Where there are power asymmetries, facilitators or independent mediators can help ensure that weaker voices are not lost. Without such attention to process, cooperative ideals give way to the same top-down dynamics they were meant to replace.

While our single-case study has limitations of context and scope, it underlines the high stakes in attempting to combine solidarity principles with competitive digital services. Future research might use comparative case methods to examine municipalities that have succeeded in launching stable cooperative platforms. Such work could reveal how variations in regulatory policy, local social capital, or partnerships with the private sector shape outcomes. There is also a need for longitudinal studies to track cooperative platforms after initial funding winds down, to see whether communities can sustain momentum once political attention shifts elsewhere. Finally, deeper theoretical inquiry might focus on refining the "partner state" concept for digital initiatives, identifying precisely how public agencies can devolve power while maintaining oversight.

In summary, City P's experiment shows that even in a context where cooperatives are legally protected and digital infrastructure is advanced, building a true "digital solidarity economy" is an immense challenge. The project's breakdown highlights the perennial complexities of aligning social mission with practical governance and advanced technology. It also reminds us that simply labeling a venture as community-owned does not guarantee genuine participation if the daily realities of technological usability, resource constraints, and power imbalances remain unaddressed. Lessons drawn from this case can guide policymakers, practitioners, and researchers as they strive to create inclusive, trusted, and technically robust digital platforms that genuinely serve the public good. By applying these insights—ensuring authentic stakeholder empowerment, high-level technical competence, long-term state support without overreach, and clear transparency mechanisms—future projects may stand a better chance of fulfilling the aspirations of platform cooperativism in a way that resonates with and benefits local communities.